\documentclass[aps,prd,twocolumn,groupedaddress,showpacs,nofootinbib]{revtex4}

\usepackage{graphicx}
\usepackage{dcolumn}
\usepackage{bm}

\usepackage{amsmath, amsthm}
\def\ben{\begin{enumerate}} \def\een{\end{enumerate}}
\def\beq{\begin{equation}} \def\eeq{\end{equation}}
\def\beqn{\begin{equation*}} \def\eeqn{\end{equation*}}
\def\bea{\begin{eqnarray}} \def\eea{\end{eqnarray}}
\def\ba{\begin{array}} \def\ea{\end{array}}
\def\beann{\begin{eqnarray*}} \def\eeann{\end{eqnarray*}}
\def\beasn{\begin{sneqnarray}} \def\eeasn{\end{sneqnarray}}

\begin{document}

\title{Gravitational observables, intrinsic coordinates, and canonical maps}

\author{J. M. Pons}
\thanks{email: {\tt pons@ecm.ub.es}}
\affiliation{Departament d'Estructura i Constituents de la Mat\`eria and Institut de
Ci\`encies del Cosmos, Universitat de Barcelona,
Diagonal 647, 08028 Barcelona, Catalonia, Spain}
\author{D. C. Salisbury}
\thanks{E-mail:\tt{dsalisbury@austincollege.edu}}
\affiliation{Max Planck Institut f\"ur Wissenschaftsgeschichte,
Boltzmannstrasse 22,
14195 Berlin, Germany\\ Department of Physics,
Austin College, Sherman, Texas 75090-4440, USA}
\author{K. A. Sundermeyer}
\thanks{E-mail:\tt{ksun@gmx.de}}
\affiliation{Freie Universit\"at Berlin, Fachbereich Physik,
Institute for Theoretical Physics, Arnimallee 14, 14195 Berlin,
Germany}

\pacs{4.20.Fy, 4.60.Ds.}

\begin{abstract}
It is well known that in a generally covariant gravitational theory
the choice of spacetime scalars as coordinates yields phase-space
observables (or ``invariants''). However their relation to the
symmetry group of diffeomorphism transformations has remained
obscure. In a symmetry-inspired approach we construct invariants out
of canonically induced active gauge transformations. These invariants may 
be intepreted as the full set of dynamical variables evaluated in the 
intrinsic coordinate system.
The functional invariants can explicitly be written as a Taylor
expansion in the coordinates of any observer, and the coefficients
have a physical and geometrical interpretation. Surprisingly, all
invariants can be obtained as limits of a family of canonical
transformations. This permits a short (again geometric) proof that
all invariants, including the lapse and shift, satisfy Poisson
brackets that are equal to the invariants of their corresponding
Dirac brackets.

\end{abstract}

\maketitle

Precisely as a consequence of the general covariance of Einstein's
theory of gravity, spatial and temporal coordinates possess no {\it
a priori} physical significance. Consequently it has long been
recognized that physically observable quantities in a Hamiltonian
formulation of general relativity must be insensitive to
transformations from any one set of coordinates to any other set.
However, procedures that have been proposed in the literature to
date for constructing observable quantities in generally covariant
theories suffer a major flaw. The procedures do not establish a link
to the full group of underlying diffeomorphism symmetries, including
transformations that alter the time coordinate. Consequently,
although the resulting phase space functionals that have been
proposed are indeed invariant under transformations generated by the
first class constraints of the theory, it has not been clear what
relationship these constraints bear to the original symmetry group.

Initial steps in elucidating this relationship were undertaken by
Pons, Salisbury, and Shepley\cite{pss1997pr}. It was pointed out
that there is  a realizeable symmetry group in the Hamiltonian
formulation of general relativity that is induced by the original
diffeomorphism group. Yet every transformation of the metric field
and its time derivatives that results from the original group is
faithfully represented by the induced symmetry group. Pons and
Salisbury\cite{ps04}  suggested that this group could be employed to
perform a transformation to intrinsic coordinates, defined through
gauge conditions in terms of appropriately selected spacetime scalar
fields. The resulting invariants are then simply correlations
between the values of these scalars and the remaining phase space
variables of the theory\cite{bergmann61}. Furthermore, invariants are nothing other
than the full set of dynamical variables evaluated in the intrinsic
coordinate system, including the conventionally discarded lapse and
shift fields.

We follow up here with an explicit general symmetry-based
construction of invariants, based on the canonical implementation of
intrinsic gauge conditions.  We are able to display the invariants
as power series in the coordinates. Since time and spatial position
are intrinsically defined, they are themselves invariant. The
coefficients of the series terms are also all invariants and {\it
ipso facto} constants of the motion. We prove in addition that the
invariants so obtained satisfy the Dirac algebra that follows from
the intrinsic coordinate-dependent gauge fixing. Finally, we show
that all invariants may be obtained as a limit of a family of
canonical transformations, thus affording a potentially useful
approximation scheme.

The diffeomorphism-induced canonical symmetry group is the group of
transformations of metric components and their canonical conjugates
that is projectable under the Legendre map from
configuration/velocity space to phase space. The corresponding
infinitesimal coordinate transformations have a compulsory
dependence on the lapse and shift metric components, and due to the
group property, a spatially non-local dependence on the spatial
components of the metric must also be admitted. Explicitly the
coordinate transformations that produce  projectable variations in
the metric are of the form \beq x'^\mu = x^\mu - n^\mu(x) \xi^0
(g_{ab};x) - \delta^\mu_a \xi^a (g_{bc};x), \label{infintrans} \eeq
 where Greek indices are in the range $0, \cdots, 3$, and Latin indices have the spatial range $1, \cdots, 3$. The ``descriptors'' $\xi^\mu$ are arbitrary functionals. The normal to the fixed time hypersurface is expressed in terms of the lapse field $N$ and shift fields $N^a$ as $n^\mu = \left( N^{-1}, - N^{-1} N^a \right)$. The phase space generator of these active transformations is
\begin{equation}
   G_{{\xi }}(t) =    P_{\mu} \dot\xi^{\mu} + ( {\cal H}_{\mu}
+ N^{\rho} C^{\nu}_{\mu \rho} P_{\nu}) \xi^{\mu} . \label{thegen}
\end{equation}
The $ P_{\mu} $ are the momenta conjugate to $N^\mu$, and are primary constraints. The $ {\cal H}_{\mu}$ are secondary constraints satisfying the first class closed Poisson bracket condition $\left\{ {\cal H}_{\mu}, {\cal H}_{\nu} \right\} =  C^{\mu}_{\nu \rho} {\cal H}_{\rho} $. We employ the convention that repeated indices represent both a sum and an integration over spatial coordinates. As opposed to (\ref{thegen}), the Hamiltonian generator of time evolution is $H =      {\cal H}_{\mu}N^{\mu} + P_{\mu} \Lambda^\mu$, where the $\Lambda^\mu$ are, excepting for the condition that $\Lambda^0 > 0$, arbitrary spacetime functions.

As a first step towards the explicit form of the functional
invariants we impose an intrinsic
coordinate-dependent gauge condition of the form $\chi^{1 \mu} := x^\mu
- X^\mu(x) = 0$, where the $X^\mu$ are spacetime scalar functions of the
canonical fields. Our task is then to find the canonical
transformation that moves the field variables to that location on
the gauge orbit where the gauge conditions are satisfied.  Once we
are in possession of this finite transformation we may employ it to
transform all the remaining fields. These actively transformed
fields are the invariants that we seek.

We note that preservation of the gauge conditions under temporal
evolution leads to  additional constraints $\chi^{2 \mu}:=\delta^\mu_0
- {\cal A}^\mu_{\ \nu}  N^\nu  \approx 0$. where we have introduced
the matrix ${\cal A}^\mu_{\ \nu} := \left\{ X^\mu, {\cal H}_{\nu}
\right\}$. Following the lead of Henneaux and
Teitelboim\cite{henneauxt94}, further exploited by
Dittrich\cite{dittrich07} and Thiemann\cite{thiemann06}, we find it
convenient to work with linear combinations $\overline{\zeta}_{(i) \nu}$
of the first class constraints $\zeta_{(1) \mu} :=  {\cal H}_{\mu},
\zeta_{(2) \mu} :=P_\mu$  having the property
that $\left\{ \chi^{(i) \mu}, \overline{\zeta}_{(j) \nu} \right\} \approx
-\delta^i_j \delta^\mu_\nu $.  For this purpose we need the
inverse of the matrix $\left\{ \chi^{(i) \mu}, \zeta_{(j) \nu} \right\}$.
The appropriate linear combinations are therefore $\overline{P}_\mu
= {\cal B}^\nu_{\ \mu}P_\nu$, where ${\cal B}^\alpha_{\ \beta}$ is
the inverse of ${\cal A}^\mu_{\ \nu}$, and $ {\overline {\cal
H}}_\nu = {\cal B}^\rho_{\ \nu}\left( {\cal H}_\rho - {\cal
B}^\mu_{\ \lambda}N^\sigma \{A^\lambda_{\sigma},\,{\cal H}_\rho\}
P_\mu \right)$. This new basis exists in general only locally.
 The gauge generator expressed in terms of this new basis is \bea
   G_{{\xi }}(t)& = & {\cal A}^\nu_{\ \mu}{\overline P}_\nu \dot\xi^\mu +
   \left( {\cal A}^\nu_{\ \mu}{\overline {\cal
H}}_\nu+ {\cal B}^\lambda_{\ \rho}N^\sigma {\cal A}^\rho_{\ \sigma},\,{\cal H}_\mu\}P_\lambda \right.\nonumber\\
&+& \left. N^\sigma C^\lambda_{\mu\sigma}P_\lambda \right)\xi^\mu\,.
\label{G} \eea
Letting $\xi = {\cal B}^\mu_{\ \sigma}{\overline\xi}^\sigma$ we find
$$\dot\xi^\mu =
 \frac{d}{d\,t}({\cal B}^\mu_{\ \sigma}{\overline\xi}^\sigma)=\{{\cal B}^\mu_{\ \sigma},\,
 N^\lambda {\cal H}_\lambda\} {\overline\xi}^\sigma +
 {\cal B}^\mu_{\ \sigma} \dot{\overline\xi^\sigma}\,,
$$
and substituting into (\ref{G}) we obtain
\beq
   G_{\overline{\xi }}(t) = {\overline P}_\nu\dot{\overline\xi^\nu} + {\overline {\cal
H}}_\nu{\overline\xi^\nu}
+ P_\mu N^\sigma{\cal S}^\mu_{\rho\sigma}{\overline\xi^\rho}\,, \label{G1bar}
\eeq
where
$$
{\cal S}^\mu_{\rho\sigma}= \{{\cal B}^\mu_{\ \rho},\,{\cal H}_\sigma\} + {\cal B}^\nu_{\ \rho}{\cal B}^\mu_{\ \gamma}\{{\cal A}^\gamma_{\ \sigma},\,{\cal H}_\nu\} + {\cal B}^\nu_{\ \rho}C^\mu_{\nu\sigma}\,.
$$
It turns out that $ {\cal S}^\mu_{\rho\sigma} = -{\cal B}^\mu_{\
\beta}{\cal B}^\alpha_{\ \rho}
\{X^\beta,\,C^\lambda_{\alpha\sigma}\}{\cal H}_\lambda$\cite{pss08}.
Thus since the last term in (\ref{G1bar}) is quadratic in the
constraints, it can be discarded ``on-shell" (on the first-class
constraint hypersurface). We obtain the following simple form for the
gauge generator, \beq G_{\overline{\xi }}(t) ={\overline
P}_\nu\dot{\overline\xi^\nu} + {\overline {\cal
H}}_\nu{\overline\xi^\nu}. \eeq

The finite active gauge transformation of any dynamical field $\Phi$
takes the form \beq exp\left(  \{ - , \, G_{\overline{\xi}}\}
\right) {\Phi} = \Phi + \left\{ \Phi , G_{\overline{\xi}}\right\} +
\frac{1}{2} \left\{ \left\{\Phi, G_{\overline{\xi}}\right\},
G_{\overline{\xi}} \right\} + \cdots \label{phi} \eeq Since by
assumption the scalars $X^\mu$ do not depend on the lapse and shift
the descriptors for the finite gauge transformation that transforms
$X^\mu$ to $x^\mu$ is easily seen to satisfy, on shell, \beq x^\mu =
X^\mu + \overline{\xi}^\mu, \eeq where $\overline{\xi}^\mu  := {\cal
A}^\nu_\sigma\xi^\sigma$ is considered a function of the spacetime
coordinates and we have made use of the fact that the Poisson
brackets ${\overline {\cal H}}_\mu$ with themselves vanish
``strongly'', i.e., they are proportional to terms at least
quadratic in the ${\overline {\cal H}}_\nu$. The descriptors
$\overline{\xi}^\mu = \chi^{(1)\mu}$ and $\dot{\overline{\xi}}^\mu =
\chi^{(2) \mu}$ may therefore be substituted into (\ref{phi}), after
the Poisson brackets have been computed, to obtain all invariant
functionals ${\cal I}_\Phi$ associated with the fields $\Phi$.

Although the following results hold for all fields \cite{pss08} we
focus here on fields other than the lapse and shift. Then the
explicit expressions for the invariants are \bea {\cal
I}_{\Phi} &\approx& \Phi + \chi^{(1) \mu}\{ \Phi , \,
 \overline{\cal H}_{\mu}\} + \frac{1}{2!}\chi^{(1) \mu}\chi^{(1) \nu}\{\{ \Phi , \,
 \overline{\cal H}_{\mu}\}, \,\overline{\cal H}_{\nu}\} + \cdots \nonumber\\
&=:&\sum_{n=0}^{\infty} \frac{1}{n!}(\chi^{(1)})^n \{ \Phi , \,
 \overline{\cal H}}\}_{(n)\,, \label{expansion}
\eea where $\approx$ is the symbol for Dirac's weak equality, that
is, an equality which holds ``on-shell". The time rate of change of
these invariant functionals satisfies
 \bea \frac{d}{d\,t} {\cal
I}_{\Phi} &=& \frac{\partial }{\partial \,t}{\cal I}_{\Phi} +
\{{\cal I}_{\Phi},\,N^\mu{\cal H_\mu}\}
\approx \frac{\partial }{\partial \,t}{\cal I}_{\Phi}
\approx \{ \Phi^A , \,
 \overline{\cal H}_{0}\}\nonumber \\
&+&\frac{1}{2} \left(\chi^{(1)\ 0}\chi^{(1)\ \nu}\{\{ \Phi^A , \,
 \overline{\cal H}_{0}\}, \,\overline{\cal H}_{\nu}\} \right.\nonumber \\ &+&\left. \chi^{(1)\ 0}\chi^{(1)\ \nu}\{\{ \Phi^A , \,
 \overline{\cal H}_{\nu}\}, \,\overline{\cal H}_{0}\} \right) +\cdots \nonumber \\
&\approx& \sum_{n=0}^{\infty} \frac{1}{n!}\,(\chi^{(1)})^n \{\{ \Phi , \,
 \overline{\cal H}_0\}, \,\overline{\cal H}\}_{(n)} ={\cal I}_{\{ \Phi , \,
 \overline{\cal H}_0\}}\,, \label{timeder}
\eea where in the third line we used the strong vanishing of the
Poisson brackets of the $\bar{\cal H}$. But recognizing that the
invariant fields are simply the fields at the gauge-fixed point
$p_{{}_{\!G}}$ on the gauge orbit, we note that ${\cal I}_{\{ \Phi ,
\,\overline{\cal H}_0\}}= {\{ \Phi , \,\overline{\cal
H}_0\}}_{p_{{}_{\!G}}} = \{ \Phi , \,N^\mu {\cal
H}_\mu\}_{p_{{}_{\!G}}}$. Therefore the invariants satisfy the
equations of motion of the gauge fixed fields. We note also that
repeated time derivatives of the invariant fields yield the  simple
expression $\frac{\partial^n}{\partial\,t^n} {\cal
I}_{\Phi}\approx{\cal I}_{\{ \Phi , \,\overline{\cal H}_0\}_{(n)}}$,
and that we may therefore express the invariants as follows as a
Taylor series in $t$: \beq {\cal I}_{\Phi}\approx
\sum_{n=0}^{\infty} \frac{t^n}{n!}\,  {\cal I}_{\{ \Phi ,
\,\overline{\cal H}_0\}_{(n)}}{}_{\big|_{t= 0}}\,. \label{evolv}
\eeq This expression displays in a striking manner the notion of
``evolving constant of the motion'' introduced originally by
Rovelli\cite{rovelli90}. We stress however, that the evolution we
obtain here is nothing other than the evolution determined by
Einstein's equations in the gauge-fixed coordinate system.  We point
out in addition that according to Batlle {\it et al}
\cite{Batlle:1987ek}
 the potentially infinite set of invariants
 ${\cal I}_{\{ \Phi , \,\overline{\cal H}_0\}_{(n)}}$ will correspond to Noether symmetries. In generic general relativity the
 conserved quantities must necessarily be spatially non-local, as pointed out by Torre\cite{torre93}.

The invariant functionals ${\cal I}_{\{ \Phi , \,\overline{\cal
H}_0\}_{(n)}}{}_{\big|_{t= 0}}$ have an immediate physical
interpretation: They are the n'th time derivative of $\Phi$
evaluated at time zero.  A similar result holds for spatial
derivatives, as we now show. But first we point out
 that the $X^\mu$ are to be spacetime scalars, i.e.,
under the projectable infinitesimal coordinate transformations
(\ref{infintrans}) we require that $\left\{X^\mu, \xi^\nu {\cal
H}_\nu \right\} = X^\mu_{, \nu} \epsilon^\nu$.  It follows that
${\cal A}^\mu_{\ a} = X^\mu_{,a}$, but there are no restrictions on
${\cal A}^\mu_{\ 0}$. At the gauge fixed point on the gauge orbit we
find therefore that ${\cal A}^\mu_{\ a} = \delta^\mu_a$, and  ${\cal
B}^\mu_{\ \nu} =N^\mu \delta^0_\nu + \delta^\mu_a \delta^a_\nu$.
Paralleling for the spatial coordinates the computation in
(\ref{timeder}) we find that $\frac{d}{d\,x^a} {\cal I}_{\Phi}
\approx {\cal I}_{\{ \Phi , \, \overline{\cal H}_a\}}$. But again we
recognize that the invariant associated with $\{ \Phi , \,
\overline{\cal H}_a \}$ is this quantity evaluated where the gauge
condition is satisfied. At this point $\overline{\cal H}_a = {\cal
B}^\mu_{\ a} {\cal H}_\mu = {\cal H}_a$, confirming that this
invariant is indeed nothing other than the spatial derivative. In
addition we may expand in powers of the spatial coordinate $x^a$,
resulting in the interpretation of the associated invariant
coefficients of n'th power of $x^a$ as the n'th spatial derivative
evaluated at $x^a = 0$. All observers end up with the same explicit
functions of their own coordinates as does the observer at the gauge
fixed location. This essentially is the physical content of the
invariants ${\cal I}_{\Phi}$.

The map $\Phi\to {\cal I}_{\Phi}$ cannot be canonical because it sends all the gauge equivalent configurations
to the same configuration, the one satisfying the gauge fixing conditions. But surprisingly,  it can be understood as a limit of a one-parameter family of canonical maps. Let us consider the functionals, obtained from canonical maps,
${\cal K}_{\Phi}^{(\Lambda)} = exp\left(  \{ - ,\Lambda\,\chi^{(1) \nu} \overline{\cal H}_{\nu}\} \right)\Phi$. One can prove that, on shell, \cite{pss08}
\beq {\cal K}_{\Phi}^{(\Lambda)}
\approx \sum_{n=0}^{\infty}\frac{1}{n!}(1-e^{-\Lambda})^n(\chi^1)^n \{ \Phi , \,
 \overline{\cal H}\}_{(n)}=:\tilde{\cal K}_{\Phi}^{(\Lambda)}\,,
\label{theK}
\eeq
It is remarkable that the same functional ${\cal K}_{\Phi}^{(\Lambda)}$ exhibits two different power series expansions:
the first is just its definition and is in terms of the parameter $\Lambda$, the second, as in (\ref{theK}),
is in terms of $(1-e^{-\Lambda})$ and is only valid on shell. The important result is that we recover the invariants,
$$\lim_{\Lambda\rightarrow\infty}{\cal K}_{\Phi}^{(\Lambda)} \approx {\cal I}_{\Phi}\,.
$$
Another result of interest is
\beq\{{\cal K}_{\Phi}^{(\Lambda)},\,\overline{\cal H}_\nu\}
\approx e^{-\Lambda} {\cal K}_{\{\Phi,\,\overline{\cal H}_\nu\}}^{(\Lambda)}\,,\label{kh}
\eeq
which again shows, in taking the $\Lambda\to\infty$ limit, the gauge invariance of ${\cal I}_{\Phi}$.

These functionals ${\cal K}_{\Phi}^{(\Lambda)}$ can be used to yield
a simple proof that the Poisson brackets of the invariants are
simply the invariants associated with the Dirac brackets of the
fields. The proof proceeds as follows
for variables other than the lapse and shift, but can  be
generalized to include them. The computation of the first order off
shell terms of ${\cal K}_{\Phi}^{(\Lambda)}$ gives \beq {\cal
K}_{\Phi}^{(\Lambda)}=\tilde{\cal K}_{\Phi}^{(\Lambda)}
+e^{\Lambda}\gamma_{{}_{\!\Phi}}^\mu\overline{\cal H}_\mu + {\cal
O}(2)\,, \eeq

with $$\gamma_{{}_{\!\Phi}}^\mu :=  (1-e^{-\Lambda})  \{\Phi,\,\chi^{1 \mu}\}
+ \frac{(1-e^{-\Lambda})^2}{2} \{\Phi,\,\overline{\cal H}_\nu\} \{\chi^{1 \nu},\,\chi^{1 \mu}\}\,.$$
The functionals $\gamma_{{}_{\!\Phi}}$ carry the information of the lowest order off shell terms for
${\cal K}^{(\Lambda)}_{\Phi}$. ${\cal O}(2)$ represents terms which are second order of the type $\alpha^{\mu\nu}\overline{\cal H}_\mu\overline{\cal H}_\nu$ or $\beta_{\mu}^{\nu}D^{1 \mu}\overline{\cal H}_\nu$ .

Now we can compute
\bea
\{\tilde{\cal K}^{(\Lambda)}_{\Phi^A},\,\tilde{\cal K}^{(\Lambda)}_{\Phi^B}\} &=&{\cal K}^{(\Lambda)}_{\{\Phi^A,\,\Phi^B\}}
- e^{\Lambda}\{{\cal K}^{(\Lambda)}_{\Phi^A},\,\overline{\cal H}_\mu\}\gamma_{{}_{\!\Phi^B}}^\mu\nonumber \\
&-& e^{\Lambda}\gamma_{{}_{\!\Phi^A}}^\mu\{\overline{\cal H}_\mu,\,{\cal K}^{(\Lambda)}_{\Phi^B}\}
+ {\cal O}(\overline{\cal H},\,\chi^1)\,,\nonumber
\eea
and when we go on shell ($\overline{\cal H}\approx 0$), using (\ref{kh}),
\bea
\{\tilde{\cal K}^{(\Lambda)}_{\Phi^A},\,\tilde{\cal K}^{(\Lambda)}_{\Phi^B}\}&\approx& {\cal K}^{(\Lambda)}_{\{\Phi^A,\,\Phi^B\}}
-  {\cal K}^{(\Lambda)}_{\{\Phi^A,\,\overline{\cal H}_\mu\}}\gamma_{{}_{\!\Phi^B}}^\mu \nonumber \\
&-& \gamma_{{}_{\!\Phi^A}}^\mu {\cal K}^{(\Lambda)}_{\{\overline{\cal H}_\mu,\,\Phi^B\}}+ {\cal O}(\chi^1)\,.\nonumber
\eea

Next we can take the limit $\Lambda\rightarrow\infty$ and obtain
\bea
\{{\cal I}_{\Phi^A},\,{\cal I}_{\Phi^B}\}&\approx& {\cal I}_{\{\Phi^A,\,\Phi^B\}}
-  {\cal I}_{\{\Phi^A,\,\overline{\cal H}_\mu\}}(\lim_{\Lambda\to\infty}{\gamma}_{{}_{\!\Phi^B}}^\mu)
\nonumber \\ &-& (\lim_{\Lambda\to\infty}\gamma_{{}_{\!\Phi^A}}^\mu) {\cal I}_{\{\overline{\cal H}_\mu,\,\Phi^B\}}
+ {\cal O}(\chi^1)\,,\nonumber
\eea
which explicitly shows the role played by the off shell terms of
${\cal K}^{(\Lambda)}_{\Phi}$ in the computation of $\{{\cal I}_{\Phi^A},\,{\cal I}_{\Phi^B}\}$. These terms
will gently conspire to bring the Dirac bracket on the stage. Indeed,
sending $\chi^1\rightarrow 0$, that is, examining the configurations at the location $p_{{}_{\!G}}$ (on the gauge orbit) where the gauge fixing constraints are satisfied, we obtain
\bea
&&\{{\cal I}_{\Phi^A},\,{\cal I}_{\Phi^B}\}{}_{\big|_{p_{{}_{\!G}}}}=
\{ \Phi^A,\,\Phi^B\} \nonumber \\
&-& \{\Phi^A,\,\overline{\cal H}_\mu\}\Big(   \{\Phi^B,\,\chi^{1 \mu}\}
+ \frac{1}{2} \{\chi^{1 \mu},\,\chi^{1 \nu}\}\{\overline{\cal H}_\nu,\,\Phi^B\} \Big) \nonumber \\
&-& \Big(   \{\Phi^A\,\chi\} + \frac{1}{2} \{\Phi^A\,\overline{\cal
H}_\nu\} \{\chi^{1 \nu},\,\chi^{1 \mu}\}\Big) \{\overline{\cal
H}_\mu,\,\Phi^B\} \nonumber \\&=&\{
\Phi^A,\,\Phi^B\}^*{}_{\big|_{p_{{}_{\!G}}}} ={\cal
I}_{\{\Phi^A,\,\Phi^B\}^*}{}_{\big|_{p_{{}_{\!G}}}} \,. \eea Now,
gauge transformations will move this result at $p_{{}_{\!G}}$ to any
other location on the gauge orbit. Since these  gauge
transformations are canonical and both the Poisson bracket structure
and the invariants are preserved by them, we can drop the
restriction $p_{{}_{\!G}}$ in the above result and conclude that
\beq \{{\cal I}_{\Phi^A},\,{\cal I}_{\Phi^B}\} \approx {\cal
I}_{\{\Phi^A,\,\Phi^B\}^*}\,. \label{DB3} \eeq This result has been
obtained previously by Thiemann\cite{thiemann06} in a much more
laborious manner, and not including lapse and shift.

We illustrate some of these classical ideas with a spatially
homogeneous isotropic cosmological model that has proven useful in
loop quantum gravitational approaches to quantum
cosmology\cite{bojowald07}. The spacetime increment squared in this
model is $ds^2 = -N^2 dt^2 + a^2 \left(dx^2 + dy^2+ dz^2\right)$ where the
lapse $N$ and expansion factor $a$ depend on the coordinate time
$t$. The material source is taken to be a massless spacetime scalar
field $\phi(t)$ It is convenient to make a change of variables in
this model  to new variables $n,u,v$ satisfying $N =: 2 n/9$,
$a=:\left(2 \pi/3\right)^{1/3} \exp\left((u+v)/3\right)$, and $\phi
=: \left(27 \pi \right)^{-1/2} \ln \left( u/v \right)$. Then the
Lagrangian becomes simply $L = - \dot u \dot v/n$ resulting in the
primary constraints $H := -p_u p_v \approx 0$ and $p_n \approx 0$.
The generator of time evolution is $n H + \lambda p_n$. Choosing
$p_v \approx 0$ we find that if we fix the gauge by $\chi^{(1)}=
t-v=0$, this model possesses the curious property that the only
physical variable that changes in time is time itself! This property
must be distinguished from the conventional notion of ``frozen
time'' in which it was understood that time did not exist. Of
course, by other gauge choices the remaining variables actually
undergo non-trivial time evolution. One can simply perform a passive
coordinate transformation on the invariant variables that we obtain
presently.

With the choice $X^0 = v$ we find ${\cal A} := \{ v, H \} = - p_u$,
so ${\cal B} = -1/p_u$, resulting in the constraints $\overline{H} =
p_v$ and $\overline{p}_n = - p_n/p_u$. The gauge generator is
therefore $G_{\overline{\xi}} = -\dot{\overline\xi} p_n/p_u +
\overline\xi p_v $. All canonical variables except $n$ commute with
$G_{\overline{\xi}}$, so ${\cal I}_u = u$, ${\cal I}_{p_u} = p_u$.
Furthermore  $\{n, G_{\overline{\xi}} \} = -\dot{\overline\xi^\nu}
/p_u  $ has a vanishing Poisson bracket with $G_{\overline{\xi}}$,
and when evaluated at $\dot{\overline\xi^\nu} = 1+p_u n = \chi^{(2)}$
results in ${\cal I}_n = n + \{n, G_{\overline{\xi}} \} = -1/p_u$.
Note that, as is generally the case, this last invariant can be
obtained through the equivalent passive coordinate transformation
from an arbitrary time $t$ to $t=v$. Representing the functional
form in the intrinsic coordinate system by $\hat n$, we have $\hat
n(v(t)) = n(t) \frac{dt}{dv} = n(t)/\frac{dv}{dt} = -1/p_u$.
Illustrating the Noether symmetry generated by the constant of
motion $u$, we find $\{u, H \} = H/p_u$, and therefore the
infinitesimal rigid symmetry generator is $\epsilon Q = \epsilon (u
+ n p_n/p_u)$. The  sole resulting variation is $\delta n = \epsilon
n/p_u = -  \epsilon/\dot u$. We confirm that the corresponding
variation of the Lagrangian is a total time derivative: $\delta L =
- \epsilon \dot v$.

Let us next consider the limit of the canonical transformation
engendered by $ \Lambda \left( -\chi^{(2)} p_n/p_u + \chi^{(1)} p_v
\right) = \Lambda \left( - p_n/p_u + n p_n + (t - v)p_v \right)$.
We find, for example,  ${\cal K}^{(\Lambda)}_n = -1/p_u + \exp(-
\Lambda) \left( 1/p_u + n \right)$, confirming that  ${\cal K}^{(\Lambda)}_n
\rightarrow {\cal I}_n$ as $\Lambda \rightarrow \infty$. Finally we
note that since the only non-vanishing elements of the inverse of
the matrix of Poisson brackets of constraints and gauge conditions
$C^1 := H$, $C^2 := \chi^{(1)}$, $C^3 := p_n$, and $C^4 := \chi^{(2)}$ are
$M^{-1}_{12} = M^{-1}_{34} = 1/p_u$ and $M^{-1}_{23} = 1/p_u^2$, the
Dirac bracket
$$
\{n, u \}^* = - \{n, p_n \} M^{-1}_{34} \{ n p_u, u \} = n/p_u,
$$
and therefore ${\cal I}_{\{n, u \}^*} = -1/p_u^2$. As expected, this is equal to the ordinary Poisson bracket $\{ {\cal I}_n, {\cal I}_u \}$.

\end{document}